\documentclass[twocolumn,aps,showpacs,floatfix,prc]{revtex4}
\usepackage[dvips]{epsfig}
\begin{document}

\title{ Yields of neutron-rich nuclei by actinide photofission in giant dipole resonance region }
\author{ Debasis Bhowmick$^1$, Debasis Atta$^2$, D. N. Basu$^3$ and Alok Chakrabarti$^4$}

\affiliation{ Variable  Energy  Cyclotron  Centre, 1/AF Bidhan Nagar, Kolkata 700 064, India }

\email[E-mail 1: ]{dbhowmick@vecc.gov.in}
\email[E-mail 2: ]{datta@vecc.gov.in}
\email[E-mail 3: ]{dnb@vecc.gov.in}
\email[E-mail 4: ]{alok@vecc.gov.in}

\date{\today }

\begin{abstract}

    Photofission of actinides is studied in the region of nuclear excitation energies that covers the entire giant dipole resonance (GDR) region. A comparative analysis of the behavior of the symmetric and asymmetric modes of photon induced fission as a function of the average excitation energy of the fissioning nucleus is performed. The mass distributions of $^{238}$U photofission fragments are obtained at the endpoint bremsstrahlung energy of 29.1 MeV which corresponds to mean photon energy of 13.7$\pm$0.3 MeV that coincides with GDR peak for $^{238}$U photofission. The integrated yield of $^{238}$U photofission as well as charge distribution of photofission products are calculated and its role in the production of neutron-rich nuclei and their exoticity is explored.

\end{abstract}

\maketitle

\noindent
\section{ Introduction }
\label{section1}

    Photonuclear reactions along with neutron-induced fission, has been studied for several years. New reliable studies on the mass distributions of fission fragments have become necessary in order to produce neutron rich ion beams \cite{TRIUMF12,Di99} (especially in the vicinity of $^{78}$Ni) which, in turn, will open new vistas in the study of nuclei far away from the valley of stability. The use of the energetic electrons is a promising tool to get intense neutron-rich ion beams. Photofission of Uranium is a very powerful mechanism to produce such radioactive ion beams (RIB). Although the photofission cross section at giant dipole resonance (GDR) energy for $^{238}$U is about an order of magnitude lower than for the 40 MeV neutron induced fission, still it is advantageous because the electrons/$\gamma$-photons conversion efficiency is much more significant than that for the deuterons/neutrons. For producing neutron-rich radioactive ion beam by the photofission method, nuclei are excited by photons covering the peak of the GDR where the energetic beam of incident electrons of $\sim$30 MeV can be slowed down in a tungsten (W) converter or directly in the target (U) itself, generating bremsstrahlung photons \cite{Es80} which can induce fission.  

    In the present study, the total photoabsorption cross section at energies covering the GDR region are calculated while the mass distributions are analyzed on the basis of the multimode-fission model \cite{Br90,Du01} and the charge distribution of photofission products are estimated. The mass distribution is interpreted as a sum of contributions of various fission modes: specifically, a symmetric mode (SM) and two asymmetric modes (ASMI and ASMII) associated with an enhanced yield of fission products. The average excitation energy for such a nucleus is used for the predictions of the multimode-fission model. A calculation of the contributions from various fission modes to the mass distribution of photofission fragments was performed in \cite{De08} at two accelerator energies. Without quantitative calculations, the existence of contributions from various fission modes was highlighted in \cite{Po93}. In the present work, we perform a simultaneous analysis and a comparison of the behavior of the symmetric and asymmetric modes of photofission induced by bremsstrahlung photons in the region of excitation energies of the $^{238}$U at the endpoint bremsstrahlung energy of 29.1 MeV which corresponds to mean photon energy of 13.7$\pm$0.3 MeV \cite{Is14} that coincides with GDR peak for $^{238}$U photofission. The results obtained in this way are compared with the predictions of the multimode-fission model for the dependence of individual fission modes on the excitation energy of the fissioning nucleus. The integrated yield of $^{238}$U photofission and charge distribution of photofission products are calculated. The role of photofission mass yield and its charge distribution in the production of neutron-rich nuclei are explored.
    
\noindent
\section{ The GDR photoabsorption and fission }
\label{section2}    
  
      In the hydrodynamic theory of photonuclear reactions, the giant dipole resonance consists of Lorentz line for spherical nuclei \cite{Go48,St50}, corresponding to the absorption of photons which induce oscillations of the neutron and proton fluids in the nucleus against each other, and the superposition of two such lines for statically deformed spheroidal nuclei \cite{Ok56,Da58}, corresponding to oscillations along each of the non-degenerate axes of the spheroid. The lower-energy line corresponds to oscillations along the longer axis and the higher-energy line along the shorter, since the absorption frequency decreases with increasing nuclear dimensions. Therefore, the semiclassical theory of the interaction between photons and nuclei entails that the shape of a fundamental resonance in the absorption cross section is given by \cite{St50,Da58} 

\vspace{-0.0cm}
\begin{equation}
\sigma_a^{GDR}(E_{\gamma})= {\bf \huge\Sigma}_{i=1}^2 \frac{\sigma_i}{1+{\Big [} \frac{(E_{\gamma}^2-E_i^2)}{E_{\gamma} \Gamma_i}{\Big ]} ^2}
\label{seqn1}
\vspace{-0.0cm}
\end{equation}
\noindent
where $\sigma_i$, $E_i$ and $\Gamma_i$ are the peak cross section, resonance energy and full width at half maximum, respectively. We find that like the photoabsorption cross sections, the photofission cross sections can also be described quite well by Eq.(1). The list of parameters $\sigma_i$, $E_i$ and $\Gamma_i$ for $i=1,2$ extracted by fitting experimental data \cite{Ve73,Ca80} are provided in Table-I for both photoabsorption as well as photofission cross sections. It may be observed from Table-I that for photoabsorption and photofission cross sections there are little changes in parameters $E_i$ and $\Gamma_i$ while $\sigma_i$ decides the difference. Therefore, we find by fitting data of all the eight actinide nuclei that ratios $R_1=\frac{(\sigma_1)_f}{(\sigma_1)_a}$ and $R_2=\frac{(\sigma_2)_f}{(\sigma_2)_a}$ for gamma absorption and subsequent fission scale as $c_1\zeta_f^2$ and $c_2\zeta_f^2$ respectively with $c_1=0.9978$, $c_2=1.4775$ where $\zeta_f$ \cite{Se58} is given by      
  
\vspace{-0.0cm}
\begin{equation}
 \zeta_f=10.9-0.319Z^2/A
\label{seqn2}
\vspace{-0.0cm}
\end{equation}
\noindent
with $Z, A$ being the charge, mass numbers of the target nucleus implying $R_{i=1,2}$ depends quadratically as $a_{2i}f^2+a_{1i}f+a_{0i}$ on the fissility $f=Z^2/A$. In Fig.-1, the measured photoabsorption and photofission cross sections (full circles) for $^{238}$U  as functions of incident photon energy are compared with the predictions (solid lines) of the hydrodynamic theory of photonuclear reactions for the giant dipole resonance region that consists of Lorentz line shapes for spherical nuclei. The dotted line represents photofission cross sections obtained by using ratio method described above. It is apparent from Fig.-1 that the ratio method predictions for photofission are almost as good as the Lorentz line shape fitting while the evaporation-fission process of the compound nucleus largely overestimates \cite{Mu10} the photofission cross sections.   
  
\begin{table*}[htbp]
\vspace{0.0cm}
\caption{\label{tab:table1} The extracted values of the parameters $\sigma_i$, $E_i$ and $\Gamma_i$, with $i=1,2$, obtained by fitting the experimental data \cite{Ve73,Ca80} for the photoabsorption and the photofission cross sections. }
\begin{tabular}{|c|c|c|c|c|c|c|}
\hline
Nuclei & $E_1$ & $\sigma_1$&$\Gamma_1$& $E_2$ & $\sigma_2$&$\Gamma_2$ \\
&[MeV] & [mb] & [MeV] & [MeV] & [mb] & [MeV]        \\  
\hline

$^{232}$Th&0.111569E+02&0.253310E+03&0.368781E+01&0.140337E+02&0.332045E+03&0.484684E+01 \\
Photoabsorption          &$\pm$0.208178E-01&$\pm$0.427193E+01&$\pm$0.671125E-01&$\pm$0.234962E-01&$\pm$0.379444E+01&$\pm$0.627879E-01 \\
&&&&&&\\
$^{232}$Th&0.108239E+02&0.148941E+02&0.154977E+01&0.142884E+02&0.386921E+02&0.339084E+01 \\
Photofission          &$\pm$0.570896E-01&$\pm$0.101739E+01&$\pm$0.196303E+00&$\pm$0.407358E-01&$\pm$0.984476E+00&$\pm$0.165864E+00\\ \hline

$^{238}$U&0.110359E+02&0.304084E+03&0.244344E+01&0.139696E+02&0.384504E+03&0.424934E+01 \\
Photoabsorption          
&$\pm$0.757291E-02&$\pm$0.191041E+01&$\pm$0.211513E-01&$\pm$ 0.105020E-01&$\pm$0.151267E+01&$\pm$0.215363E-01\\
&&&&&&\\
$^{238}$U&0.108646E+02&0.509458E+02&0.235096E+01&0.142606E+02&0.139631E+03&0.431961E+01\\
Photofission          &$\pm$0.345136E-01&$\pm$0.141568E+01&$\pm$0.104248E+00&$\pm$0.241084E-01&$\pm$0.119854E+01&$\pm$0.610462E-01\\ \hline

$^{236}$U&0.109999E+02&0.349761E+03&0.200208E+01&0.139997E+02&0.444907E+03&0.400213E+01\\
Photoabsorption          &$\pm$0.469940E-01&$\pm$0.163054E+02&$\pm$0.155638E+00&$\pm$0.848482E-01&$\pm$0.134267E+02&$\pm$0.194748E+00\\
&&&&&&\\
$^{236}$U&0.108181E+02&0.766137E+02&0.211832E+01&0.142140E+02&0.233508E+03&0.482840E+01\\
Photofission  
&$\pm$0.498453E-01&$\pm$0.395287E+01&$\pm$0.177282E+00&$\pm$0.482288E-01&$\pm$0.328896E+01&$\pm$0.125866E+00\\ \hline

$^{235}$U&0.109997E+02&0.349758E+03&0.107362E+01&0.140000E+02&0.445123E+03&0.400294E+01\\
Photoabsorption          
&$\pm$0.558415E-01&$\pm$0.355239E+02&$\pm$0.166314E+00&$\pm$0.743161E-01&$\pm$0.160659E+02&$\pm$0.225709E+00\\
&&&&&&\\
$^{235}$U&0.109097E+02&0.955815E+02&0.198927E+01&0.138757E+02&0.380470E+03&0.421280E+01\\
Photofission          
&$\pm$0.592284E-01&$\pm$0.513518E+01&$\pm$0.188900E+00&$\pm$0.287584E-01&$\pm$0365952E+01&$\pm$0.704117E-01\\ \hline

$^{234}$U&0.110543E+02&0.210986E+03&0.108276E+01&0.131269E+02&0.446181E+03&0.606000E+01\\
Photoabsorption          
&$\pm$0.328566E-01&$\pm$0.123333E+02&$\pm$0.138675E+00&$\pm$0.575317E-01&$\pm$0.719500E+01&$\pm$0.882948E-01\\
&&&&&&\\
$^{234}$U&0.110819E+02&0.163781E+03&0.281607E+01&0.145927E+02&0.348757E+03&0.407321E+01\\
Photofission          
&$\pm$0.330435E-01&$\pm$0.347119E+01&$\pm$0.879498E-01&$\pm$0.210564E-01&$\pm$0.321142E+01&$\pm$0.654639E-01\\ \hline

$^{233}$U&0.110510E+02&0.239833E+03&0.181537E+01&0.139116E+02&0.437354E+03&0.543458E+01 \\
Photoabsorption          
&$\pm$0.183016E-01&$\pm$0.551579E+01&$\pm$0.743190E-01&$\pm$0.312191E-01&$\pm$0.381128E+01&$\pm$0.564126E-01\\
&&&&&&\\
$^{233}$U&0.110246E+02&0.142698E+03&0.173286E+01&0.138589E+02&0.407635E+03&0.408954E+01\\
Photofission          
&$\pm$0.675373E-01&$\pm$0.113748E+02&$\pm$0.226983E+00&$\pm$0.582266E-01&$\pm$0.836806E+01&$\pm$0.134014E+00\\ \hline

$^{237}$Np&0.110277E+02&0.246408E+03&0.279883E+01&0.141381E+02&0.392209E+03&0.516690E+01 \\
Photoabsorption          
&$\pm$0.160029E-01&$\pm$0.436347E+01&$\pm$0.550226E-01&$\pm$0.204200E-01&$\pm$0.241868E+01&$\pm$0.738509E-01 \\
&&&&&&\\
$^{237}$Np&0.109213E+02&0.158435E+03&0.255789E+01&0.143494E+02&0.261391E+03&0.410446E+01\\
Photofission          
&$\pm$0.579898E-02&$\pm$0.696476E+00&$\pm$0.205505E-01&$\pm$0.523555E-02&$\pm$0.537081E+00&$\pm$0.195039E-01\\ \hline

$^{239}$Pu&0.111363E+02&0.310118E+03&0.225973E+01&0.135343E+02&0.376374E+03&0.472413E+01 \\
Photoabsorption          &$\pm$0.330433E-01&$\pm$0.124836E+02&$\pm$0.117752E+00&$\pm$0.691318E-01&$\pm$0.100811E+02&$\pm$.918920E-01\\
&&&&&&\\
$^{239}$Pu&0.112990E+02&0.207497E+03&0.271930E+01&0.143349E+02&0.311287E+03&0.376570E+01\\
Photofission          
&$\pm$0.366080E-01&$\pm$0.549059E+01&$\pm$0.874060E-01&$\pm$0.347835E-01&$\pm$0.448887E+01&$\pm$0.910799E-01\\

\hline
\end{tabular} 
\vspace{0.0cm}
\end{table*}
  
\begin{figure}[htbp]
\vspace{0.0cm}
\eject\centerline{\epsfig{file=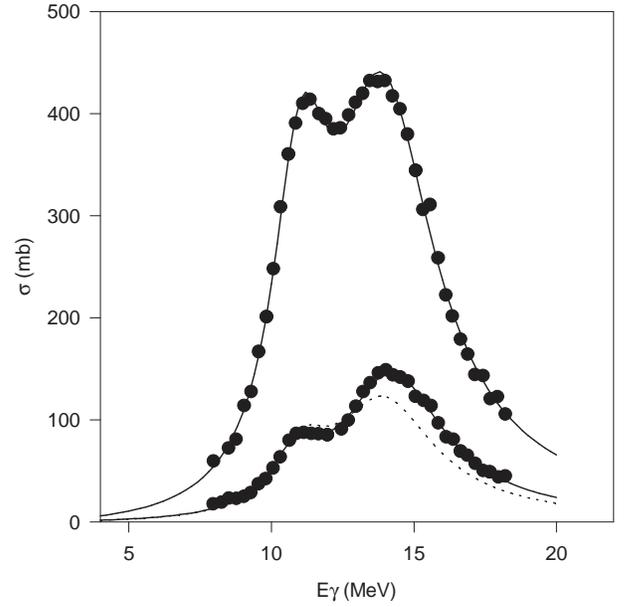,height=8cm,width=8cm}}
\caption
{Comparison of the measured photoabsorption and photofission cross sections (full circles) for $^{238}$U  as functions of incident photon energy with predictions (solid lines) of the hydrodynamic theory of photonuclear reactions for the giant dipole resonance region that consists of Lorentz line shapes for spherical nuclei. The dotted line represents photofission cross sections obtained by using ratio method. }
\label{fig1}
\vspace{0.0cm}
\end{figure}
\noindent 
    
\noindent
\section{ Mass yield distribution of photofission products }
\label{section3}

    In the multimode-fission model, the mass distribution is interpreted as a sum of the contributions
from the symmetric and asymmetric fission modes. Each fission mode corresponds to the passage through the fission barrier of specific shape. For each fission mode, the yield is described in the form of a Gaussian function. However, it is impossible to approximate the shape of the mass distributions by using only three Gaussian functions (two fission modes). For this, one needs five Gaussian functions (three fission modes). For $^{238}$U fission, the symmetric fission mode (SM) is associated $A=117$ and for the asymmetric fission modes (ASMI, ASMII) in addition to broad maxima at $A=138$ and $A=96$, the mass distribution exhibits narrower maxima in mass-number regions around $A=133$ and $A=101$. Thus the total yield of fragments whose mass number is $A$ is given by the expression
    
\begin{eqnarray}
 Y(A) =&& Y_{SM}(A) + Y_{ASMI}(A) + Y_{ASMII}(A)  \nonumber\\
 =&& C_{SM}\exp \Big[-\frac{(A-A_{SM})^2}{2\sigma^2_{SM}} \Big] \\
 && + C_{ASMI}\exp \Big[-\frac{(A-A_{SM}-D_{ASMI})^2}{2\sigma^2_{ASMI}} \Big] \nonumber\\
 && + C_{ASMI}\exp \Big[-\frac{(A-A_{SM}+D_{ASMI})^2}{2\sigma^2_{ASMI}} \Big] \nonumber\\
 && + C_{ASMII}\exp \Big[-\frac{(A-A_{SM}-D_{ASMII})^2}{2\sigma^2_{ASMII}} \Big] \nonumber\\ 
 && + C_{ASMII}\exp \Big[-\frac{(A-A_{SM}+D_{ASMII})^2}{2\sigma^2_{ASMII}} \Big]. \nonumber
\label{seqn3}
\end{eqnarray}
\noindent

    In Fig.-2, approximation by the above five Gaussian functions for the mass distribution of fragments originating from $^{238}$U photofission induced by bremsstrahlung photons whose endpoint energy is 29.1 MeV $Y(A)$ per 100 fission events is plotted and compared with experimental data \cite{Is14}. The values of $A_{SM}$, $D_{ASMI}$ and $D_{ASMII}$ are 117, 21 and 16, respectively whereas the other parameter values are $C_{SM}=0.4929$, $\sigma_{SM}=4.4732$, $C_{ASMI}=5.8959$, $\sigma_{ASMI}=5.9612$, $C_{ASMII}=2.2945$ and $\sigma_{ASMII}=1.6223$. 

\begin{figure}[htbp]
\vspace{0.0cm}
\eject\centerline{\epsfig{file=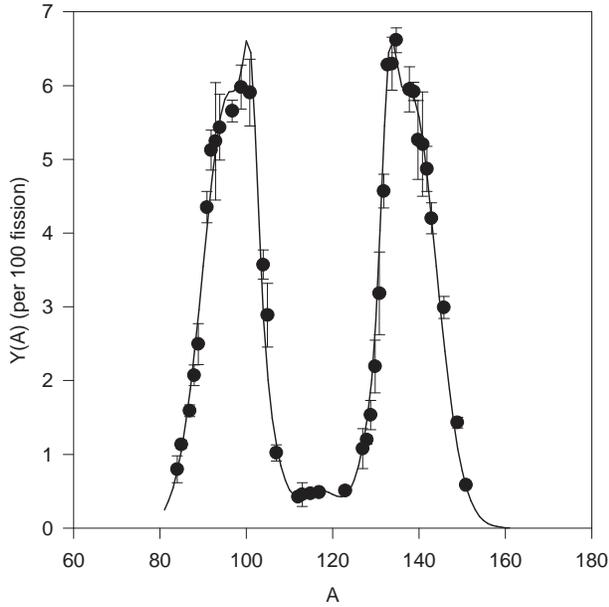,height=8cm,width=8cm}}
\caption
{Comparison of the measured mass yield distribution (full circles) for $^{238}$U photofission induced by bremsstrahlung photons whose endpoint energy is 29.1 MeV with the prediction (solid line) of the five Gaussian formula for $Y(A)$.}
\label{fig2}
\vspace{0.0cm}
\end{figure}
\noindent

\noindent
\section{ Charge distribution of photofission products }
\label{section4}

    The isobaric charge distribution of photofission products can be well simulated by a Gaussian function as
 
\begin{equation}
 Y(A,Z) = \frac{Y(A)}{\sqrt{\pi C_p}} \exp \Big[-\frac{(Z-Z_s+\Delta)^2}{C_p} \Big]
\label{seqn4}
\end{equation}
\noindent    
where  $Z_s$ represents most stable isotope of fission fragment with mass number $A$ while $\Delta$ measures the departure of the most probable isobar from the stable one. In order to deduce expression for $Z_s$, theoretically, for the most stable nucleus by keeping mass number $A$ constant while differentiating liquid drop model mass formula and setting the term $\partial M_{nucleus}(A,Z)/\partial Z\mid_A$ equal to zero as \cite{Ch06} 

\begin{equation}
 Z_s = \frac{[A+(a_cA^{2/3}/2x)]}{[(4a_{sym}/x)+(a_cA^{2/3}/x)]}
\label{seqn5}
\end{equation}
\noindent
where $x=2a_{sym}+[(m_n-m_p)/2]$, $a_c$ and $a_{sym}$ being the Coulomb and symmetry energy coefficients, respectively, whereas $m_p$ and $m_n$ are the masses of proton and neutron respectively. The values of the parameter $C_p$ which decides the dispersion and the shift parameter $\Delta$ for the most probable isotope are extracted by fitting experimental data \cite{Is14} to be 0.8 and 3.8, respectively. The fractional yield $DY(A,Z)$ is defined as the ratio of the independent yield of production of nuclei belonging to a specific mass and charge number, $Y(A,Z)$, to the total yield $Y(A)$ for the same mass number $A$, that is, $DY(A,Z) = Y(A,Z)/Y(A)$. In Fig.-3, charge distribution as fractional yield $DY(A,Z)$ is plotted for fragment mass number $A=97$ of $^{238}$U photofission. In Fig.-3, fractional yield $DY(A,Z)$ as a function of charge number $Z$ for a particular fragment mass number $A=97$ of $^{238}$U photofission is plotted.  

\begin{figure}[htbp]
\vspace{0.0cm}
\eject\centerline{\epsfig{file=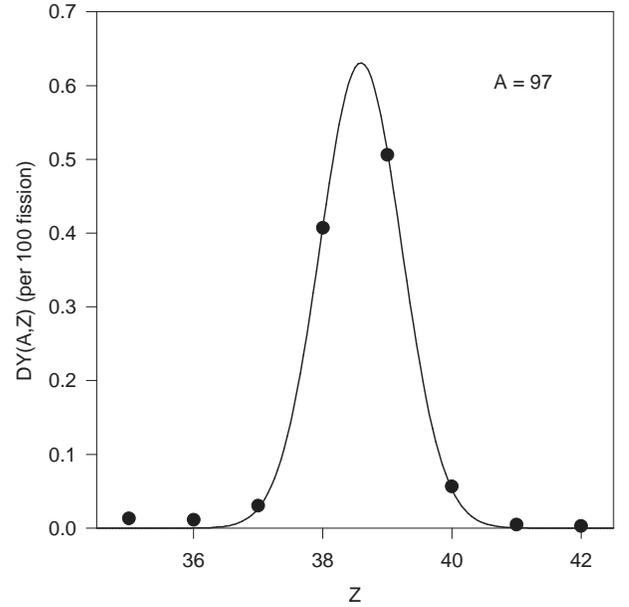,height=8cm,width=8cm}}
\caption
{Plot of fractional yield $DY(A,Z)$ as a function of charge number $Z$ for fragment mass number $A=97$ of $^{238}$U photofission. Experimental data are shown with full circles while solid line represents the Gaussian charge distribution.}
\label{fig3}
\vspace{0.0cm}
\end{figure}
\noindent
  
\noindent
\section{ Calculation and results }
\label{section5}

    The photofission cross sections $\sigma_f^{GDR}$ are calculated using Lorentz line shape Eq.(1) for gamma absorption while replacing $\sigma_i$ by the ratio method described in section-II. The production cross sections of individual fragments for $^{238}$U photofission induced by bremsstrahlung photons whose endpoint energy is 29.1 MeV are obtained by multiplying fission cross section by charge distribution which means $\sigma_f(A,Z)= \sigma_f^{GDR}.Y(A,Z)/100$. The endpoint energy of 29.1 MeV (the energy of electrons which produce bremsstrahlung gammas when stopped by a $W$ converter) is so chosen because it corresponds to the mean gamma energy of 13.7$\pm$0.3 MeV that coincides with GDR peak for $^{238}$U photofission. In Table-II, the theoretical cross sections for most probable (produced with highest production cross section) isobars with corresponding neutron and proton numbers for the fission fragments are tabulated. Corresponding atomic numbers $Z_s$ for the most stable nuclei are calculated using Eq.(5) with values for $a_c=0.71$ MeV and $a_{sym}=23.21$ MeV \cite{Ch06}. The proton and neutron drip lines are calculated theoretically from modified liquid drop model mass formula \cite{Sa02} using logic \cite{Ba04} that the nucleus from which removal of a single neutron (and any more) makes the one proton separation energy negative defines a proton drip line nucleus and the nucleus to which addition of a single neutron (and any more) makes the one neuton separation energy negative defines a neutron drip line nucleus.  

\begin{table}[h!]
\vspace{0.0cm}
\caption{\label{tab:table1} The theoretical cross sections for most probable (produced with highest production cross section) isobars with corresponding neutron and proton numbers for the fission fragments.}
\begin{tabular}{|c|c|c|c|c||c|c|c|c|c|}
\hline
A & N& Z & $Z_s$& $\sigma$ &A & N& Z & $Z_s$& $\sigma$\\
  & &  &  & [mb]&  & &  &  & [mb]         \\  
\hline

   80&   48&   32&   36&0.11562E+00&   81&   49&   32&   36&0.18627E+00\\
   82&   49&   33&   36&0.22993E+00&   83&   50&   33&   37&0.42322E+00\\
   84&   51&   33&   37&0.50203E+00&   85&   51&   34&   38&0.78573E+00\\
   86&   52&   34&   38&0.10690E+01&   87&   52&   35&   38&0.11874E+01\\
   88&   53&   35&   39&0.18556E+01&   89&   54&   35&   39&0.18815E+01\\
   90&   54&   36&   40&0.26153E+01&   91&   55&   36&   40&0.30571E+01\\
   92&   55&   37&   40&0.29817E+01&   93&   56&   37&   41&0.40232E+01\\
   94&   57&   37&   41&0.35453E+01&   95&   57&   38&   42&0.42736E+01\\
   96&   58&   38&   42&0.43749E+01&   97&   58&   39&   42&0.37144E+01\\
   98&   59&   39&   43&0.46367E+01&   99&   60&   39&   43&0.40502E+01\\
  100&   60&   40&   44&0.48037E+01&  101&   61&   40&   44&0.48132E+01\\
  102&   61&   41&   44&0.33281E+01&  103&   62&   41&   45&0.31183E+01\\
  104&   63&   41&   45&0.18741E+01&  105&   63&   42&   46&0.14313E+01\\
  106&   64&   42&   46&0.11271E+01&  107&   64&   43&   46&0.65130E+00\\
  108&   65&   43&   47&0.64657E+00&  109&   66&   43&   47&0.44646E+00\\
  110&   66&   44&   48&0.35695E+00&  111&   67&   44&   48&0.34525E+00\\
  112&   68&   44&   48&0.25037E+00&  113&   68&   45&   49&0.32626E+00\\
  114&   69&   45&   49&0.32907E+00&  115&   69&   46&   49&0.31678E+00\\
  116&   70&   46&   50&0.39368E+00&  117&   71&   46&   50&0.32803E+00\\
  118&   71&   47&   51&0.37133E+00&  119&   72&   47&   51&0.36359E+00\\
  120&   72&   48&   51&0.27508E+00&  121&   73&   48&   52&0.33305E+00\\
  122&   74&   48&   52&0.29068E+00&  123&   74&   49&   53&0.31042E+00\\
  124&   75&   49&   53&0.39884E+00&  125&   76&   49&   53&0.38461E+00\\
  126&   76&   50&   54&0.63200E+00&  127&   77&   50&   54&0.81498E+00\\
  128&   77&   51&   54&0.94162E+00&  129&   78&   51&   55&0.15553E+01\\
  130&   79&   51&   55&0.18648E+01&  131&   79&   52&   56&0.28580E+01\\
  132&   80&   52&   56&0.41599E+01&  133&   81&   52&   56&0.37146E+01\\
  134&   81&   53&   57&0.49953E+01&  135&   82&   53&   57&0.45242E+01\\
  136&   82&   54&   57&0.38439E+01&  137&   83&   54&   58&0.45866E+01\\
  138&   84&   54&   58&0.38935E+01&  139&   84&   55&   59&0.41242E+01\\
  140&   85&   55&   59&0.42549E+01&  141&   86&   55&   59&0.30171E+01\\
  142&   86&   56&   60&0.35472E+01&  143&   87&   56&   60&0.30180E+01\\
  144&   87&   57&   60&0.22401E+01&  145&   88&   57&   61&0.22881E+01\\
  146&   89&   57&   61&0.16128E+01&  147&   89&   58&   62&0.13099E+01\\
  148&   90&   58&   62&0.11106E+01&  149&   91&   58&   62&0.65162E+00\\
  150&   91&   59&   63&0.57562E+00&  151&   92&   59&   63&0.40705E+00\\
  152&   92&   60&   63&0.22370E+00&  153&   93&   60&   64&0.19076E+00\\
  154&   94&   60&   64&0.11303E+00&  155&   94&   61&   64&0.66812E-01\\
  156&   95&   61&   65&0.47827E-01&  157&   96&   61&   65&0.23853E-01\\
  158&   96&   62&   66&0.15076E-01&  159&   97&   62&   66&0.91000E-02\\
  160&   98&   62&   66&0.38368E-02&  161&   98&   63&   67&0.25778E-02\\

\hline
\end{tabular} 
\vspace{0.0cm}
\end{table}
         
\noindent
\section{ Production of neutron-rich nuclei }
\label{section6}

    Production of neutron-rich radioactive ion beam by the photofission method is by far the most advantageous because electrons/$\gamma$-photons conversion efficiency is much more significant than that for the deuterons/neutrons which matters for neutron production. Nuclei are excited by photons covering the peak of the GDR where the energetic beam of incident electrons of $\sim$30 MeV can be slowed down in a tungsten (W) converter generating bremsstrahlung photons which can induce fission. In Fig.-4, atomic number $Z$ versus neutron number $N$ are plotted for exotic nuclei produced by the photofission of $^{238}$U at the endpoint bremsstrahlung energy of 29.1 MeV which corresponds to mean photon energy of 13.7$\pm$0.3 MeV. The results of three sets of calculations are presented which correspond to (i) nuclei with largest cross sections, that is, most probable isobars with corresponding neutron and proton numbers for the fission fragments, (ii) the most neutron rich isobars subject to production cross sections $>$100 nb and (iii) the most neutron rich isobars subject to production cross sections $>$100 fb. The experimentally observed $\beta$ stable nuclei as well as the theoretical proton and neutron drip lines are also shown in Fig.-4 in order to highlight how far one can march away from the line of $\beta$-stability towards the neutron drip line by RIB using $^{238}$U photofission. It is worthwhile to mention here that at high energies \cite{Bh98} the projectile fragment separator type RIB facilities, being developed in different laboratories, could also provide the scope for producing many new exotic nuclei through fragmentation of high energy radioactive ion beams \cite{Mu91,Au95}.

\begin{figure}[htbp]
\vspace{0.0cm}
\eject\centerline{\epsfig{file=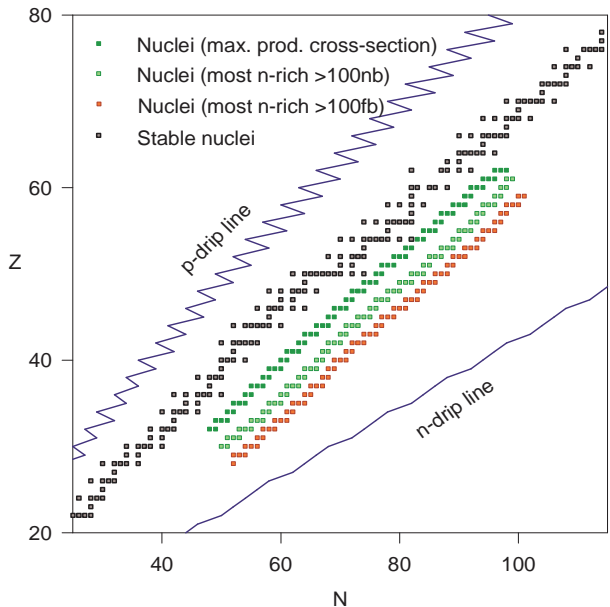,height=8cm,width=8cm}}
\caption
{Plots of atomic number $Z$ versus neutron number $N$ for exotic nuclei produced by the photofission of $^{238}$U at the endpoint bremsstrahlung energy of 29.1 MeV which corresponds to mean photon energy of 13.7$\pm$0.3 MeV.}
\label{fig4}
\vspace{0.0cm}
\end{figure}
\noindent
      
\noindent
\section{ Summary and conclusion }
\label{section7}
 
    In summary, we find that like the photoabsorption cross sections, the photofission cross sections can also be described quite well by Lorentz line shapes. The ratio method predictions for photofission are almost as good as the Lorentz line shape fitting, whereas the evaporation-fission process of the compound nucleus largely overestimates the photofission cross sections. We have performed a simultaneous analysis for the comparison of the behavior of the symmetric and asymmetric modes. The phenomenological methodology for obtaining independent and cumulative yields of isotopes produced in photofission is described. A detailed analysis of the production of each nuclear isobar via fission and the mass distributions of products originating from the photofission induced by bremsstrahlung photons, whose endpoint energy is 29.1 MeV, is provided. The endpoint energy of 29.1 MeV, which is the energy of electrons that produce bremsstrahlung gammas when stopped by a $W$ converter, is so chosen as to correspond the mean gamma energy of 13.7$\pm$0.3 MeV which coincides with GDR peak for $^{238}$U photofission. 
    
     The production of neutron-rich RIBs by the photofission method is by far the most advantageous because electrons/$\gamma$-photons conversion efficiency is much more significant than that for the deuterons/neutrons. Nuclei are excited by photons covering the peak of the GDR where the energetic beam of incident electrons of $\sim$30 MeV can be slowed down in a tungsten (W) converter generating bremsstrahlung photons which can induce fission. The results of three sets of calculations are presented which correspond to nuclei with largest cross sections, that is, most probable isobars with corresponding neutron and proton numbers for the fission fragments, the most neutron rich isobars subject to production cross sections greater than hundred nano barns and the most neutron rich isobars subject to production cross sections greater than hundred femto barns. How far one can march away from the line of $\beta$-stability towards the neutron drip line by RIB using $^{238}$U photofission is highlighted by comparison with experimentally observed $\beta$ stable nuclei as well as the theoretical proton and neutron drip lines.

\end{document}